\documentclass[12pt]{article} 
\usepackage{ol2}
\usepackage[draft]{hyperref}
\usepackage{amsmath}

\begin{document}

\title{Demonstration of temporal cloaking}


\author{Moti Fridman$^1$, Alessandro Farsi$^1$, Yoshitomo Okawachi$^1$, and Alexander L. Gaeta$^1$}

\address{
School of Applied and Engineering Physics, Cornell University, Ithaca, NY 14853, USA
}

{\bf 
Recent research has uncovered a remarkable ability to manipulate and control electromagnetic fields to produce effects such as perfect imaging and spatial cloaking\cite{Cloaking1, Cloaking2}. To achieve spatial cloaking, the index of refraction is manipulated to flow light from a probe around an object in such a way that a ``hole" in space is created, and it remains hidden\cite{Cloaking3, Cloaking4, Cloaking5, Cloaking6, Cloaking7, Valentine, carpet1, Cloaking8, Cloaking9, Cloaking10, Cloaking11, Cloaking12}. Alternatively, it may be desirable to cloak the occurrence of an event over a finite time period, and the idea of temporal cloaking was proposed in which the dispersion of the material is manipulated in time to produce a ``time hole" in the probe beam to hide the occurrence of the event from the observer\cite{Theoretical_temporal_cloaking}. This approach is based on accelerating and slowing down the front and rear parts, respectively, of the probe beam to create a well controlled temporal gap in which the event occurs so the probe beam is not modified in any way by the event. The probe beam is then restored to its original form by the reverse manipulation of the dispersion. Here we present an experimental demonstration of temporal cloaking by applying concepts from the time-space duality between diffraction and dispersive broadening\cite{agrawal}. We characterize the performance of our temporal cloak by detecting the spectral modification of a probe beam due to an optical interaction while the cloak is turned off and on and show that the event is observed when the cloak is turned off but becomes undetectable when the cloak is turned on. These results are a significant step toward the development of full spatio-temporal cloaking.
}

The detection of an object or an event is often performed by measuring a change in the properties of a light probe that interacts with the object or with elements participating in the event. The idea of spatial cloaking consists of the probe light being bent in a precise fashion to prevent it from being scattered by the object and remain hidden from an observer. This has been done typically through use of exotic materials, such as ones with a negative index of refraction or through sophisticated manipulation of the refractive index\cite{Cloaking7, Valentine, carpet1}. In analogous fashion, it could be possible to cloak an event in the time domain from an observer by manipulating the dispersion of a material such that a temporal gap is created in the probe beam, and any event that occurs within this gap does not modify the temporal/spectral properties of the probe beam and thus remains undetected\cite{Theoretical_temporal_cloaking}. This requires rapid changes in the dispersion and the recently proposed approach\cite{Theoretical_temporal_cloaking} involves the use of optical fibers that are pumped to high power levels to produce large changes in the intensity-dependent refractive index. However, at such powers, other optical processes such as stimulated Raman and Brillouin scattering could limit the ability to achieve cloaking. As such, we propose an alternative approach to create the conditions that allow for temporal cloaking in which we apply concepts from the time-space duality associated with diffraction and dispersion\cite{kolner1}.

Time-space duality represents the analogy between diffraction and dispersion that arises from the mathematical equivalence between the equations describing the diffraction of a beam of light and the one-dimensional temporal propagation of a pulse through a dispersive medium\cite{kolner1, kolner2}. Similar to a spatial lens that imparts a quadratic phase in space, a time-lens can be implemented that produces a quadratic phase shift in time\cite{TimeLens1, TimeLens2, TimeLens3}. This time-lens can, for example, magnify\cite{Foster} or compress\cite{compress} signals in time and has an equivalent of the lens law. Time-lenses can be created with an electro-optic modulator\cite{kolner2} or via a parametric nonlinear optical process such as four-wave mixing (FWM) with a chirped pump wave\cite{TimeLens1, TimeLens2}. In the latter case the signal wave is converted to an idler wave with a linear frequency shift in time (i.e., a quadratic phase in time)\cite{TimeLens3, Foster}.

To create our temporal cloaking system, we implement a split time-lens (STL) composed of two half time-lenses\cite{TimeLens1, TimeLens2, TimeLens3, Foster, compress} connected at the tips. A schematic of the temporal cloaking device is presented in Fig.~\ref{schem}, and a detailed description of these split time-lenses is presented in the supplementary methods. A continuous-wave (CW) probe beam, which is used to detect an event, is incident from the left. The split time-lens bends light to the edges of the time window rather than to the center so that after propagating through a medium with normal group-velocity dispersion (GVD), part of the probe light is delayed in time while the adjoining part is advanced. After the dispersive medium, a temporal gap is created in the probe beam so that any event that might lead to a temporal or spectral change during this temporal window will have no effect, and the event remains undetected. Finally, a medium with anomalous GVD together with a second similar split time-lens closes the hole such that neither the occurrence of the event nor the presence of the time-lenses are detected.

\begin{figure}
\centerline{\includegraphics[width=8.9cm]{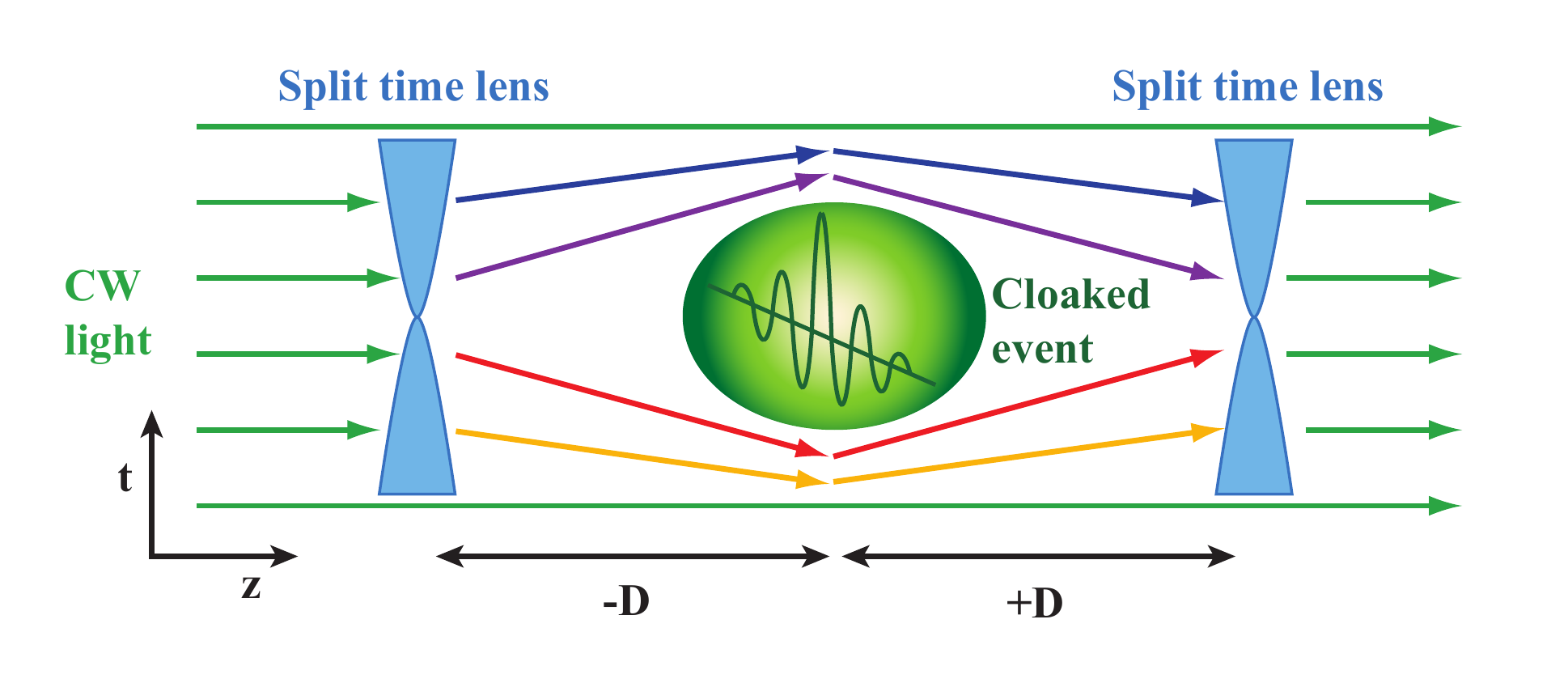}}
\caption{\label{schem} Schematics of the temporal cloak using a pair of split time-lenses (STL). The STL's are used to create a temporal ``hole" in a probe beam such that any temporal or spectral changes caused by an event within this hole do not occur. The figure is oriented such that the probe light is described by horizontal lines, and lines at different orientations represent different wavelengths. $D$ denotes the magnitude of the total negative or positive group-velocity dispersion.}
\end{figure}

\begin{figure}
\centerline{\includegraphics[width=8.9cm]{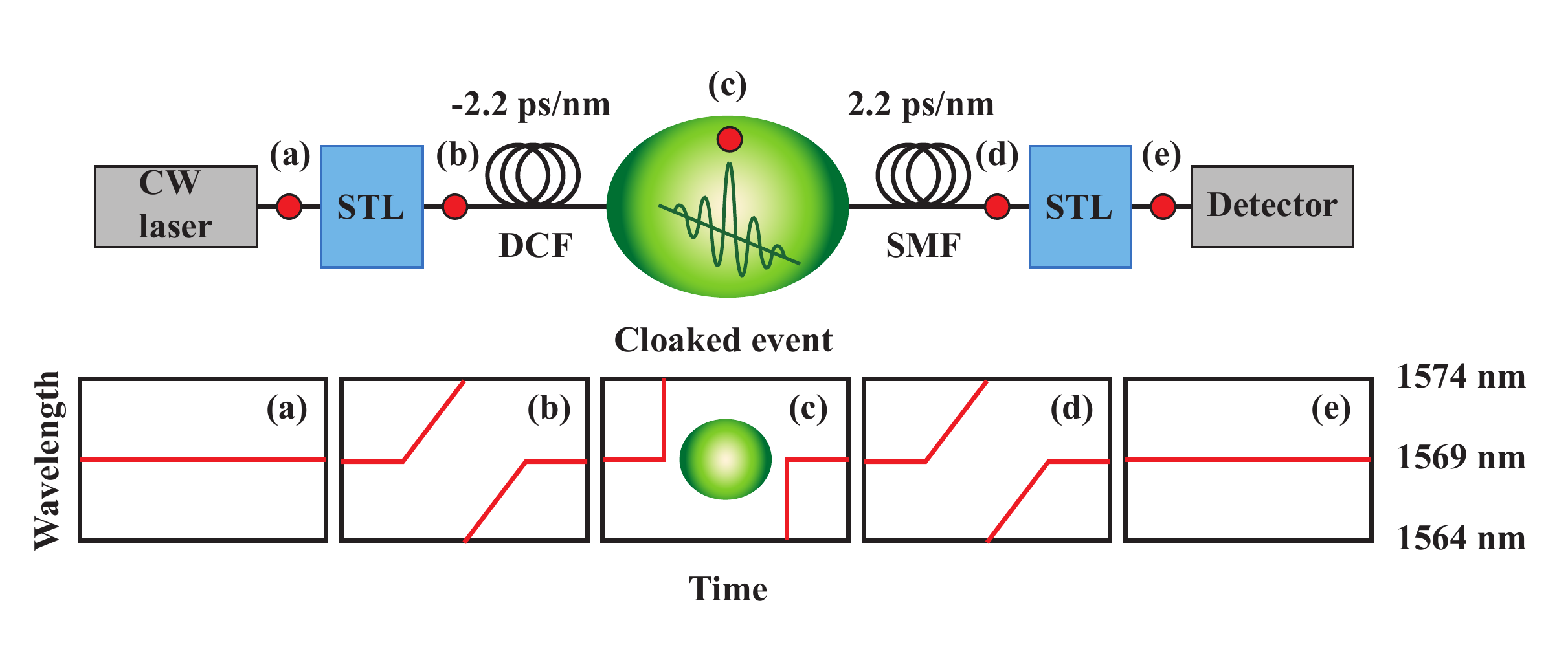}}
\caption{\label{main} Top: Experimental configuration for cloaking an event in time using two split time-lenses. Both split time-lenses are identical with the same $f$-number and are described in the supplementary methods. DCF - dispersion compensating fiber; SMF - single-mode fiber; STL - split time-lens. Bottom: The wavelength of the probe beam as a function of time. (a) Before and (b) after the first STL; (c) after propagating through the DCF; (d) before and (e) after the second STL. When both STL's are in operation the event becomes invisible.}
\end{figure}

The experimental configuration for temporal cloaking is presented in Fig.~\ref{main}. The probe beam passes through the first STL, and the wavelength of the probe beam as a function of time before and after the STL is shown in Fig.~\ref{main}(a) and Fig.~\ref{main}(b), respectively. The light then propagates through a dispersive element consisting of a dispersion-compensating fiber (DCF) such that shorter (longer) wavelengths propagate faster (slower) as compared to the initial probe beam wavelength. Thus, the wavelength distribution as a function of time shown in Fig.~\ref{main}(b) becomes that as shown in Fig.~\ref{main}(c), in which a temporal gap is opened at the focal point of the STL. The gap is synchronized such that the event occurs within this gap and thus is not sensed by the probe. The probe then propagates through a single-mode fiber (SMF), and the temporal gap is closed as shown in Fig.~\ref{main}(d). Finally, a second, identical STL restores the probe light back to its initial wavelength, as shown in Fig.~\ref{main}(e) so that the probe beam is restored to its initial state, and both the event and the presence of the time-lenses are undetected. We note that both after the time-lenses and after the event, we remove the pump waves from the system with wavelength division multiplexers (WDM).

\begin{figure}
\centerline{\includegraphics[width=8.3cm]{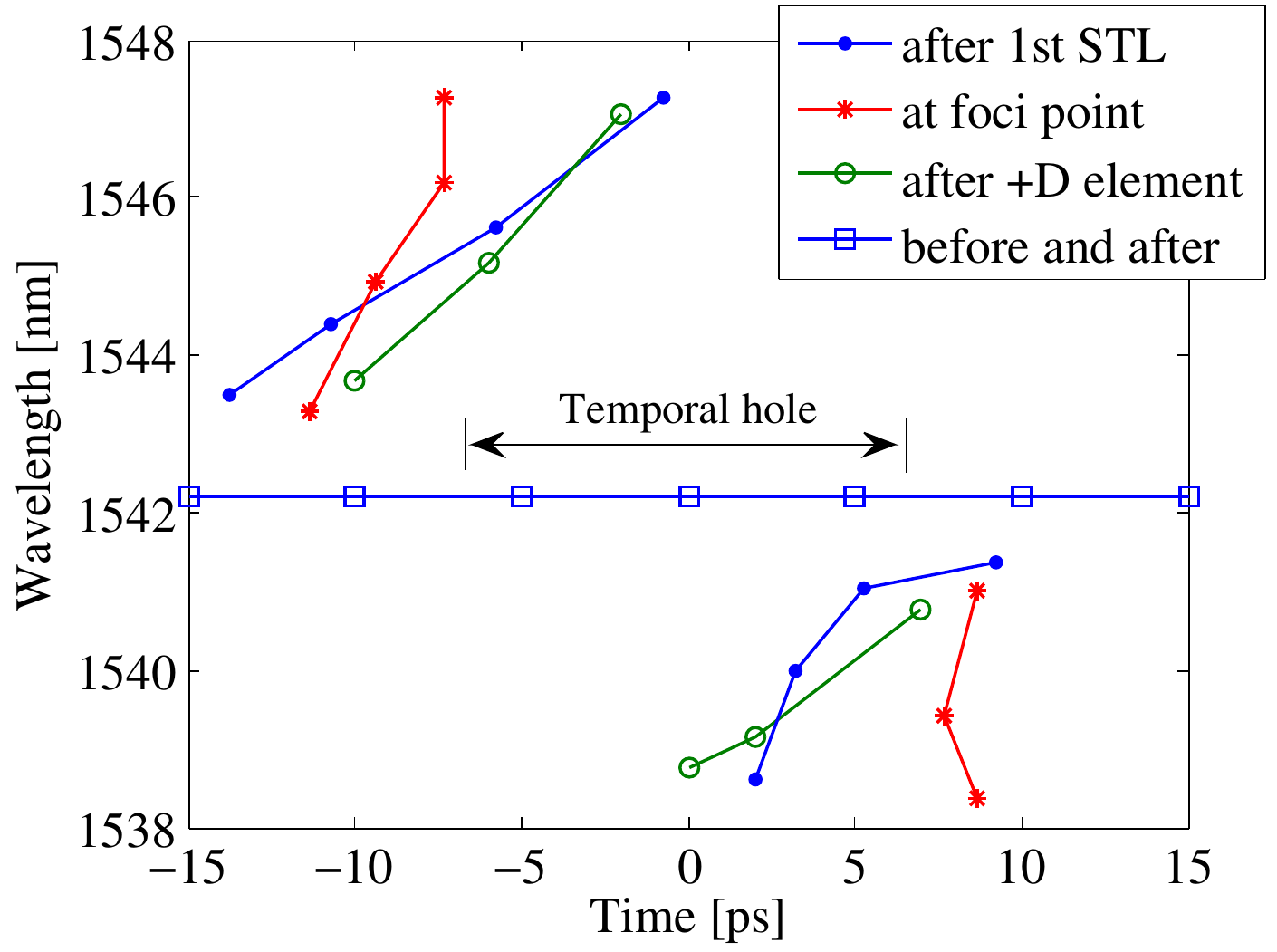}}
\caption{\label{gap} Measured wavelength distribution as a function of time at three points in the system. Dots - after the first split time-lens (STL) the wavelengths are distributed according to Fig.~\ref{main}(b); asterisks - at the focal point, after the negative dispersion a temporal gap is opened according to Fig.~\ref{main}(c); circles - after the positive dispersion before the second STL, the gap is closed and the wavelength distribution is returned to its initial state as described by Fig.~\ref{main}(b).}
\end{figure}

Figure~\ref{gap} shows experimental data illustrating how a temporal gap is created in a probe beam at 1542 nm. The wavelength of the probe as a function of time is shown after the first STL (dots), after the DCF at the focal length of the STL (asterisks), and after the SMF before the second STL (circles). After the first STL the probe acquires a frequency chirp illustrated in Fig.~\ref{main}(b). After the DCF the higher wavelengths are delayed while the lower wavelengths are advanced so that a temporal gap opens from -7.5 ps to 7.5 ps according to Fig.~\ref{main}(c). This gap is closed after the SMF, and the frequency chirp returns to that in Fig.~\ref{main}(d). Before the first STL and after the second STL we detect only the probe at its original wavelength.

\begin{figure}
\centerline{\includegraphics[width=8.3cm]{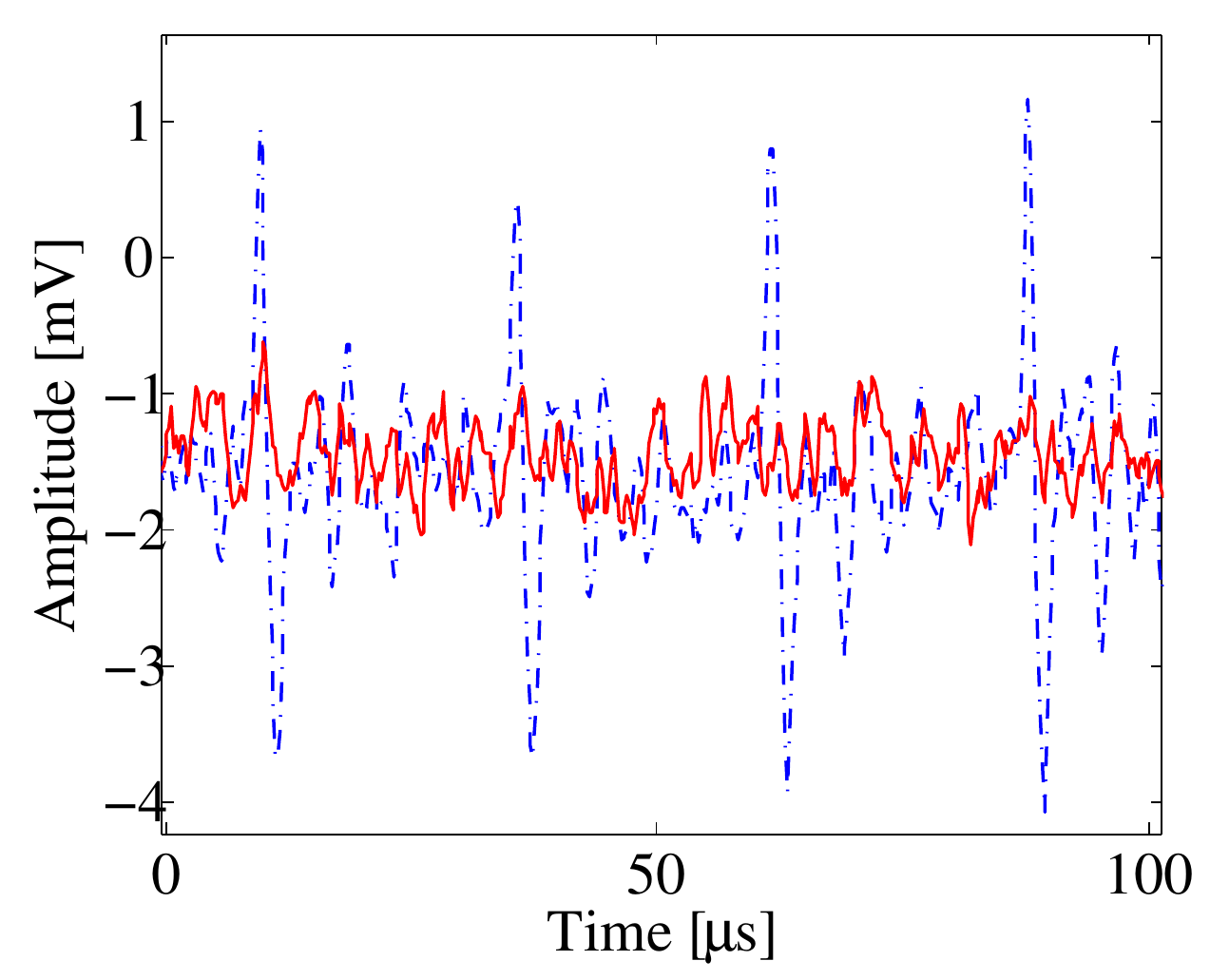}}
\caption{\label{cloakTime} Experimental results showing the detection of signal indicating that the probe beam has undergone an interaction (i.e., an event) with a short pump pulse. The events occur every 24 $\mu$s when the cloaking is turned off (dashed curve) and on (solid curve). When the cloaking is turned on, the amplitude of the signal probing the event is below the detection noise level indicating that the event has been hidden.}
\end{figure}

To demonstrate the temporal cloaking capability of this system, we create an event that results in the generation of new frequencies due to the presence of the probe beam. It consists of a nonlinear interaction of a short pump pulse with a probe beam via FWM with a repetition rate of 41 kHz. When the cloak is off, the probe beam at 1569 nm interact in a highly nonlinear fiber with a short (5 ps) pump at 1554 nm such that a frequency component is generated at 1539 nm every 24 $\mu$s. Thus, the signature of the event is the detection of the 1539 nm signal, and when the cloaking is turned off, it is clearly observed, as shown in Fig.~\ref{cloakTime} by the dashed (blue) curve. However, when the cloaking is turned on, the amplitude of the detected signal is reduced below the detection noise level, as shown in Fig.~\ref{cloakTime} [solid (red) curve] for several interaction events.

\begin{figure}
\centerline{\includegraphics[width=8.3cm]{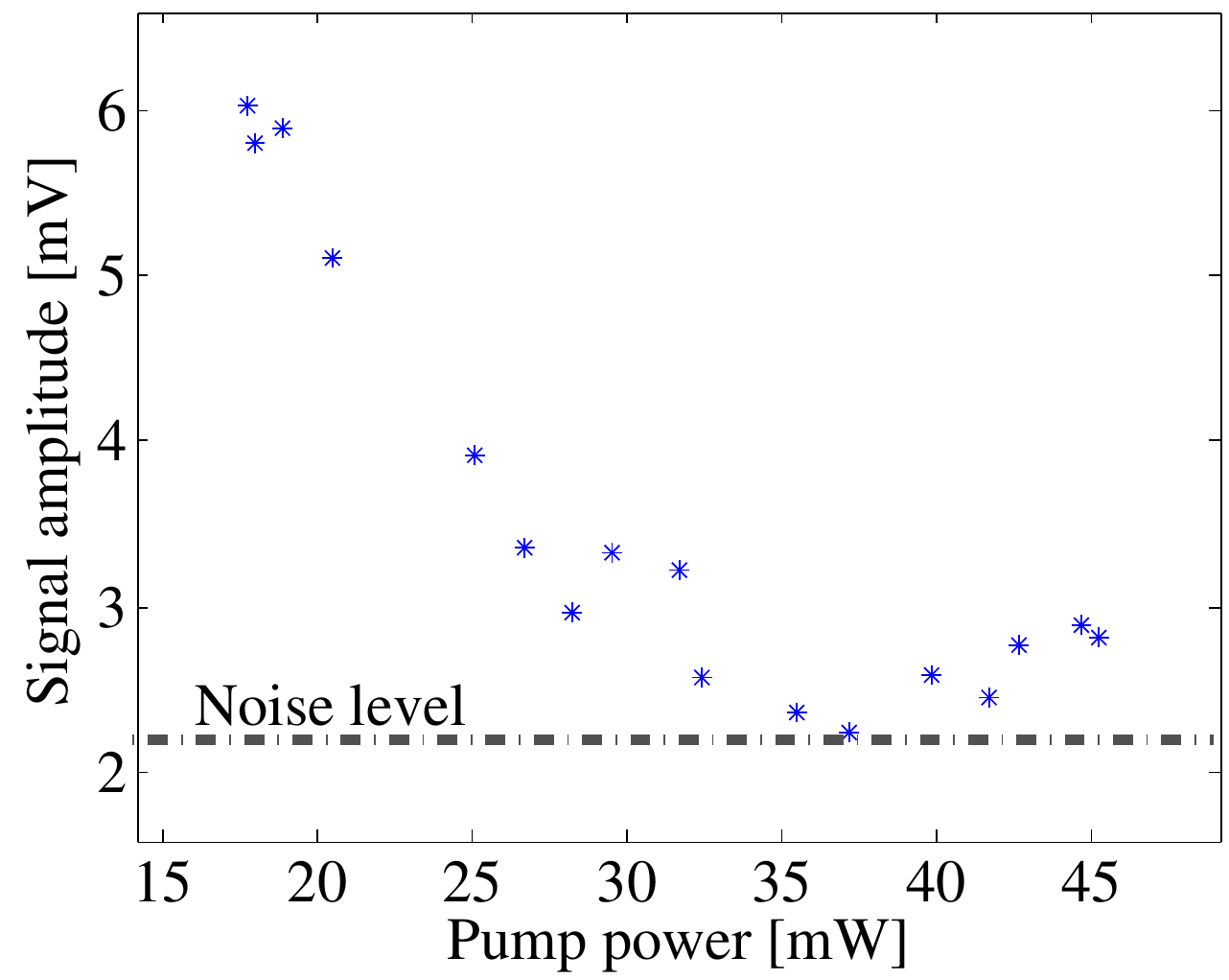}}
\caption{\label{AmpVsPump} Amplitude of the detected event as a function of the pump power of the split time-lenses.}
\end{figure}

Finally, we investigate the efficiency of the cloak as a function of the pump power of the STL's by measuring the amplitude of the detected event as a function of the pump power utilized by both STL's. The pump power of the STL's governs the amount of light in the probe beam that is shifted in frequency, so as the pump power increases, less light remains in the probe beam during the temporal gap when the event occurs (see Fig.~\ref{AmpVsPump}). When the average pump power is 17 mW, the amplitude of the detected signal is 6.1 mV and as the pump power is increased, the detected amplitude decreases until it reaches the noise level at 2.3 mV, when the pump power of the time-lens is 37 mW. Increasing the pump power of the STL's further than this point increases the amplitude due to higher pump noise.

The temporal gap can be readily widened by increasing the dispersive broadening of the pump and the dispersion $D$ between the STL's. However, as the dispersion is increased, effects due to third-order dispersion (TOD) which, if not compensated, will prevent the gap from closing completely. In our experiment, the spectral and temporal width of the pump pulses before chirping are 9 nm and 0.4 ps, respectively, which results in TOD limiting the width of the temporal gap to 110 ns, as long as the pump power is increased to efficiently deplete the probe beam completely within the gap. Nevertheless, since the TOD is proportional to $\Delta \lambda^3$ while the amount of linear frequency chirp is proportional to $\Delta \lambda^2$~\cite{agrawal}, it is possible to increase the temporal gap by resorting to narrower pump pulses and introducing more dispersion. The limitation in this case is stimulated Brillouin scattering which will limit the SMF to 50-km long and the temporal gap to 1.25 $\mu$s width.

In summary, we present the first experimental demonstration of temporal cloaking that successfully hides an event from a probe beam in the time domain. Our scheme is based on the time-space duality that uses a pair of split time-lenses. The split time-lenses utilize four-wave mixing such that dispersion manipulation is highly efficient, and can be readily adopted for other wavelengths in the electromagnetic spectrum. Our results serve as a significant step towards obtaining a complete spatio-temporal cloaking device.


\end{document}